\documentclass[twocolumn,superscriptaddress,aps]{revtex4}
\usepackage{epsfig}
\usepackage{graphicx}
\usepackage{bm}
\begin{document}
\title{A generalized unknown qubit discriminator}
\author{Bing He}
\affiliation{Department of Physics and Astronomy, Hunter College of the City University of New York, 
 695 Park Avenue, New York, NY 10021, USA}
\author{J\'{a}nos A. Bergou }
\affiliation{Department of Physics and Astronomy, Hunter College of the City University of New York, 
 695 Park Avenue, New York, NY 10021, USA}

\begin{abstract}
We discuss a state discriminator that unambiguously distinguishes between two quantum registers prepared with 
multiple copies of two unknown qubits. This device achieves the optimal performance by von Neumann measurement and general POVM in different ranges of the input preparation 
probabilities, respectively, and in the limit of very large program registers the optimal measurement reduces to 
only von Neumann measurements. 

\end{abstract}
\maketitle

The discrimination between a pair of linearly independent pure states $|\psi_{1}\rangle$ and $|\psi_{2}\rangle$ is possible only when their complete classical knowledge is known or these states carry some symmetry. Even under these conditions, their identification is achieved only with a certain probability if these states are nonorthogonal. There are strategies to achieve the optimal results. 
One is to distinguish between the states with a minimal probability of error \cite{helstrom, holevo}, and the other one of unambiguous discrimination (UD) allows the possibility of a minimal inconclusive result \cite{ivanovic,dieks,peres,js,cb}. Moreover, there is the minimax approach \cite{dariano} to deal with the situations with no information about the {\it a prior} probabilities of the states. For practical applications, one needs a device that achieves the optimal discrimination between any pair of quantum states in a versatile way---first of all, it works universally on arbitrary pairs of the target states o be discriminated and, secondly, it should perform optimally in the whole range of the {\it a prior} probabilities or preparation probabilities possibly arising in the target state preparation period.

To realize the unknown quantum states (no classical knowledge available) discrimination, we rely on the symmetry from the available multiple copies of the states we want to identify \cite{bergou,hayashi0,hayashi, he1}. Here we only consider the UD of unknown states. In Ref. [8], D'Ariano {\it et al.} propose a state discriminator in which the information about two unknown qubits is supplied in the form of a program in a programmble device \cite{dusek}. There one copy of these qubits is used as the program or reference digits. This problem was generalized to the situation of multiple copies as the reference \cite{hayashi, he1}. In Ref. \cite{hayashi}, by means of the representation theory of $U(n)$ group, Hayashi {\it et al.} obtain the optimal UD solution for the averages of two uniformly distributed unknown qudit systems prepared with $n$ copies of reference plus one copy of data, while in \cite{he1} we apply the Jordan basis method to the corresponding problem of two unknown qubit systems prepared with $n$ copies of reference plus $m$ copies of data which are prepared with arbitrary {\it a prior} probabilities. Other recent progress in unknown quantum state discriminations and their physical realization can be found in \cite{k0,he2,k1,k2,k3,k5,k6,k7,k8, k9}. Many of these studies are about the optimal UD for the averages of the systems. In this paper, we will construct a device that works on the exact input states without first taking the average of the inputs---the measurement by the device tells us which qubit it is with a non-zero success probability.

We start with the generalization of the inputs in \cite{bergou}. Here we store arbitrary $n$ copies of qubits
$|\psi_{1}\rangle$ and  $|\psi_{2}\rangle$ in the program section of the quantum registers as follows:  
\begin{eqnarray}
|\Psi_{1}^{in}\rangle & = & |\psi_{1}\rangle_{1}  |\psi_{2}\rangle_{2} \cdots |\psi_{1}\rangle_{2n-1} |\psi_{2}\rangle_{2n}
|\psi_{1}\rangle_{2n+1} \nonumber \\
|\Psi_{2}^{in}\rangle & = & |\psi_{1}\rangle_{1} |\psi_{2}\rangle_{2} \cdots |\psi_{1}\rangle_{2n-1} |\psi_{2}\rangle_{2n}
|\psi_{2}\rangle_{2n+1},
\end{eqnarray}
where the data, either $|\psi_{1}\rangle$ or $|\psi_{2}\rangle$, at the tail is appended 
to the program as a tensor product of copies of two unknown qubits. Without loss of generality we place $|\psi_{1}\rangle$ in the odd positions and $|\psi_{2}\rangle$ in the even positions of the
inputs. We need to find an optimal measurement to unambiguously distinguish between these inputs,
bearing in mind the data qubits,
 \begin{eqnarray}
|\psi_{i}\rangle=\cos(\theta_i/2)|0\rangle+\sin(\theta_i/2)
e^{i\phi_i}|1\rangle\ ,
\end{eqnarray}
where $i=1,2$, are randomly distributed on Bloch sphere with no available information 
about $\theta_i$'s and $\phi_i$'s. Generally these quantum registers are assumed to be prepared with {\it a prior}\
probabilities $\eta_1$ and $\eta_2$, respectively.
 
The optimal measurement performed by our device is generally achieved by a positive operator value measure (POVM), and we denote its element of 
unambiguosly detecting $|\Psi_{1}^{in}\rangle $ as $\Pi_1$, that of 
unambiguosly detecting $|\Psi_{2}^{in}\rangle $ as $\Pi_2$, and that corresponding to failure as $\Pi_0$. They satisfy the unit
decomposition:
\begin{eqnarray}
I=\Pi_1+\Pi_2+\Pi_0\ .
\end{eqnarray}
The probabilities of successfully identifying these two possible 
input states are 
\begin{equation}
\langle\Psi_{1}^{in}|\Pi_{1}|\Psi_{1}^{in}\rangle = p_{1} \hspace{1cm} 
\langle\Psi_{2}^{in}|\Pi_{2}|\Psi_{2}^{in}\rangle = p_{2} \ ,
\label{probs}
\end{equation}
and the condition of no error implies that
\begin{equation}
\Pi_{2}|\Psi^{in}_{1}\rangle = 0 \hspace{1cm} 
\Pi_{1}|\Psi^{in}_{2}\rangle = 0 \ .
\label{UDcond} 
\end{equation}

Without any knowledge about $|\psi_{1}\rangle$ and $|\psi_{2}\rangle$ themselves, what we can use is their symmetry, i.e. $|\Psi_{1}^{in}\rangle$ is invariant under the action of symmetric group on the odd positions and 
the tail position, while $|\Psi_{2}^{in}\rangle$ 
invariant under the corresponding action on the even positions and the tail. To construct $\Pi_1$ and $\Pi_2$
that perform the UD of the inputs, we define the following orthornormal basis:
\begin{widetext}
\begin{eqnarray}
|e_0\rangle &=& \frac{1}{\sqrt{C^0_n}}|0,0,\ldots,0\rangle \nonumber\\
|e_1\rangle &=& \frac{1}{\sqrt{C^1_n}}(|1,0,\ldots,0\rangle+|0,1,\ldots,0\rangle+\cdots+|0,0,\ldots,1\rangle)\nonumber\\
&&\cdots  \nonumber\\
|e_k\rangle&=& \frac{1}{\sqrt{C^k_n}}(\underbrace{\underbrace{|1,1,\ldots,0,0\rangle}\limits_{k's~1~in~n~digits}+
|0,1,1,\ldots,,0\rangle +\cdots+|0,0,\ldots,1,1\rangle}\limits_
{summation~of~C^k_n~terms})\nonumber\\
&&\cdots\nonumber\\
|e_n\rangle &=&\frac{1}{\sqrt{C^n_n}}|1,1,\ldots, 1\rangle\ ,
\end{eqnarray}
\end{widetext}
where $C^k_n$ is the number of ways to choose $k$ objects from a group of $n$ objects without regard to order.
In terms of these basis, the tensor product of $n$ copies of a qubit is expanded as follows:
\begin{eqnarray}
&&(\cos(\theta/2 )|0\rangle+\sin (\theta/2)
e^{i\phi}|1\rangle)^{\otimes n} \nonumber\\
&=& \sum\limits_{k=0}^{n}
\cos^{n-k}(\theta/2)\sin^{k}(\theta/2)e^{ik\phi}\sqrt{C^k_n}|e_k\rangle\ .
\end{eqnarray}
Thus we construct the following POVM elements satisfying the unambiguity (no error) in measurement:
\begin{eqnarray}
\Pi_{1} & = & c_1(I_{E,T}-P_{E,T})\otimes I_O \nonumber\\
\Pi_{2} & = & c_2(I_{O,T}-P_{O,T})\otimes I_E,
\end{eqnarray}
where $P_{E,T}=\sum\limits_{k=0}^{n+1}|e_k\rangle_{E,T}~_{E,T}\langle e_k|$ (resp. $P_{O,T}
=\sum\limits_{k=0}^{n+1}|e_k\rangle_{O,T}~_{O,T}\langle e_k|$)
is the projection operator onto the totally symmetric subspace with respect to the
$n$ even (resp. odd) positions and the tail position,
$I_{E,T}$ (resp. $I_{O,T}$) the direct product of unit matrices in
 these positions, and $c_1$, $c_2$ the non-negative parameters arising from the unequal {\it a priori} probabilities of 
$|\Psi_{1}^{in}\rangle$ and $|\Psi_{2}^{in}\rangle$. $I_O$ and $I_E$ are the 
direct products of the unit matrices in the odd and the even positions of the input
states, respectively.

Using these elements $\Pi_i$, where $i=1,2$, we find the probability of success detection as  
\begin{eqnarray}
p_i=\langle\Psi_{i}^{in}|\Pi_i|
\Psi_{i}^{in}\rangle=c_i-c_i\langle\Psi_{i}^{in}|P_{R,T}\otimes I_S|
\Psi_{i}^{in}\rangle \ ,
\end{eqnarray} 
where $R=E,O$ but $S=O,E$.
We apply two tricks to calculate the average of the projection operators more conveniently.
First, by the expansion
\begin{eqnarray}
|e_k \rangle_{R,T} &= &\sqrt{\frac{C^{k}_{n}}{C^k_{n+1}}}|e_k \rangle_{R} |0\rangle_{T}+
\sqrt{\frac{C^{k-1}_{n}}{C^k_{n+1}}}|e_{k-1}\rangle_{R} |1\rangle_{T},\nonumber\\
\end{eqnarray} 
for $k=1,2,\cdots,n$,
we rewrite the projection operators as
\begin{eqnarray}
&& P_{R,T}\otimes I_{S} \nonumber\\
&& =  \{|e_0 \rangle_R  |0\rangle_T ~_R\langle e_0|~_T\langle 0| \nonumber\\
& &+\sum\limits_{l=1}^{n}
(\sqrt{\frac{C^{l}_{n}}{C^l_{n+1}}}|e_l \rangle_{R} |0\rangle_{T}+
\sqrt{\frac{C^{l-1}_{n}}{C^l_{n+1}}}|e_{l-1}\rangle_{R} |1\rangle_{T})\nonumber\\
&&~\times(\sqrt{\frac{C^{l}_{n}}{C^l_{n+1}}}~_R\langle e_l|~_T\langle 0|+
\sqrt{\frac{C^{l-1}_{n}}{C^l_{n+1}}}~_R\langle e_{l-1}|~_T \langle 1|)\nonumber\\
& &+|e_n \rangle_R |1\rangle_T ~_R\langle e_n| ~_T\langle 1|\}\otimes I_S.\nonumber\\
\end{eqnarray} 
Second, we apply Eq. (7) to expand the input states into two parts:
\begin{widetext}
\begin{eqnarray}
|\Psi_{i}^{in}\rangle &&=\underbrace{\sum\limits_{l=0}^{n}
\cos^{n-l}(\theta_j/2)\sin^{l}(\theta_j/2)e^{il\phi_j}\sqrt{C^l_n}|e_l\rangle_R \left (\cos(\theta_i/2)|0\rangle_T+\sin(\theta_i/2)
e^{i\phi_i}|1\rangle_T \right )}\limits_{|\Psi_i^{\prime}\rangle}\nonumber\\
&&\times\underbrace{\sum\limits_{k=0}^{n}\cos^{n-k}(\theta_i/2)\sin^{k}(\theta_i/2)e^{ik\phi_i}\sqrt{C^k_n}|e_k\rangle_S}
\limits_{|\Psi^{\prime\prime}_i\rangle} \ ,
\end{eqnarray}
\end{widetext}
where $i=1,2$ but $j=2,1$.
Putting the expectation values from these parts together, we obtain
\begin{eqnarray}
&&\langle\Psi_{i}^{in}|P_{R,T}\otimes I_S|\Psi_{i}^{in}\rangle \nonumber\\
&=&\sum\limits_{k=0}^{n}\left (\frac{n-k+1}{n+1}~cos^2(\theta_i/2 )
+\frac{k+1}{n+1}~sin^2(\theta_i/2 )\right )\nonumber\\
&\times & C^{k}_{n}~ cos^{2(n-k)}(\theta_j/2)~sin^{2k}(\theta_j/2)\nonumber\\
&+&\sum\limits_{k=1}^{n}~\frac{2k}{n+1}~C^k_n ~cos^{2(n-k)}(\theta_j/2)sin^{2k}(\theta_j/2) ~cot(\theta_j/2)\nonumber\\
&\times & cos(\theta_i/2)~sin(\theta_i/2) cos(\phi_i-\phi_j)\ .
\end{eqnarray}
The total success probability $P$ to discriminate between the two unknown states, assuming $|\Psi_{1}^{in}\rangle $ is 
produced with
 {\it a priori} probability $\eta_1$ and $|\Psi_{2}^{in}\rangle $ with $\eta_2$, is given by
\begin{eqnarray}
P&=&\eta_1 p_1+\eta_2 p_2\ .
\end{eqnarray}
Since we have no knowledge about the data states, the performance of the device performs is indicated by the average 
\begin{eqnarray}
\overline{P}&=&\frac{1}{(4\pi)^2}\prod_{j=1}^{2}\int_{0}^{2\pi}d\phi_{j}\int_{0}^{\pi}d\theta_{j}\sin
\theta_{j}(\eta_1\langle\Psi_{1}^{in}|\Pi_1|\Psi_{1}^{in}\rangle\nonumber\\
&+& \eta_2 \langle\Psi_{2}^{in}|\Pi_2|\Psi_{2}^{in}\rangle) \nonumber\\
\nonumber \\
&=& (\eta_1 c_1+\eta_2 c_2)\frac{n}{2(n+1)}\ 
\end{eqnarray}
of the success probability.
We want to maximize this expression subject to the constraint that $\Pi_0$ is a positive operator.

Let $H$ be the Hilbert space of the input states given in Eq.(1), which, as it is seen from Eq.(10), is spanned by 
the orthornormal basis
$\{|e_l\rangle_O |e_m\rangle_E |0\rangle_T, |e_l\rangle_O |e_m\rangle_E |1\rangle_T\}$, where $l,m=0,1,\cdots,n$.
To have a nice matrix representation of the inconclusive operator,
\begin{eqnarray}
&& \Pi_0 =I-\Pi_1-\Pi_2\nonumber\\
&=& (1-c_1-c_2)I+c_1P_{E,T}\otimes I_O+c_2P_{O,T}\otimes I_E\ ,
\end{eqnarray}
we apply the following orthogonal transformations
\begin{eqnarray}
|\eta_{lm}\rangle&=&\sqrt{\frac{C^{m}_{n}}{C^m_{n+1}}}|e_l\rangle_{O} |e_m \rangle_{E} |0\rangle_{T}\nonumber\\
&+&
\sqrt{\frac{C^{m-1}_{n}}{C^m_{n+1}}}|e_l\rangle_{O} |e_{m-1}\rangle_{E} |1\rangle_{T}\nonumber\\
|\chi_{lm}\rangle&=& \sqrt{\frac{C^{m-1}_{n}}{C^m_{n+1}}}|e_l\rangle_{O}  |e_m \rangle_{E} |0\rangle_{T}\nonumber\\
&-& \sqrt{\frac{C^{m}_{n}}{C^m_{n+1}}}|e_l\rangle_{O} |e_{m-1}\rangle_{E} |1\rangle_{T},
\end{eqnarray}
where $l=0,1,\cdots,n$, and $m=1,2,\cdots,n$. The transformed vectors $|\eta_{lm}\rangle$ and $|\chi_{lm}\rangle$ satisfy 
$\langle\eta_{lm}|\chi_{l'm'}\rangle=0$, 
$\langle\eta_{lm}|\eta_{l'm'}\rangle=\delta_{l,l'}\delta_{m,m'}$, 
and $\langle\chi_{lm}|\chi_{l'm'}\rangle=\delta_{l,l'}\delta_{m,m'}$.
If we put them together with the untransformed $|e_l\rangle_O |e_0\rangle_E |0\rangle_T$ and $|e_l\rangle_O |e_n\rangle_E |1\rangle_T$, 
for $l=0,1,\cdots,n$, all these unit vectors will form the orthornormal basis, 
$\{|\eta_{lm}\rangle, |\chi_{lm}\rangle, |e_l\rangle_O |e_0\rangle_E |0\rangle_T, 
|e_l\rangle_O |e_n\rangle_E |1\rangle_T\}$,
of $H$. With this basis, the operator $\Pi_0$ is represented by the direct sum of two series of block diagonal matrices:
\begin{eqnarray}
\Pi_0=\Pi_0^{(1)}\oplus \Pi_0^{(2)},
\end{eqnarray}
where
\begin{eqnarray}
\Pi_0^{(1)}&=&\left(\begin{array}{ccccc}J_0 & & & &\\
 &J_1 & & &\\
 & & J_2 & &\\
 & & & \ddots & \\
 & & & & J_n\\
\end{array}\right), \nonumber\\
\Pi_0^{(2)}&=&\left(\begin{array}{ccccc}K_0 & & & &\\
 &K_1 & & &\\
 & & K_2 & &\\
 & & & \ddots & \\
 & & & & K_n\\
\end{array}\right) .
\end{eqnarray}
These block diagonal matrices are generated regularly; $J_0$ and $K_0$ are $1\times 1$ matrix $1$,
and $J_1$ is a $3\times 3$ matrix:
\begin{eqnarray}
J_1=\left(\begin{array}{ccc}
1-\frac{1}{n+1}c_2&\frac{\sqrt{n}}{(n+1)^{3/2}}c_2& -\frac{n}{(n+1)^{3/2}}c_2\\
\frac{\sqrt{n}}{(n+1)^{3/2}}c_2&1-\frac{n}{(n+1)^2}c_2&\frac{n^{3/2}}{(n+1)^2}c_2\\
-\frac{n}{(n+1)^{3/2}}c_2&\frac{n^{3/2}}{(n+1)^2}c_2&1-c_1-\frac{n^2}{(1+n)^2}c_2\\
\end{array}\right),~~~~~
\end{eqnarray}
and then the size of them grows up with the general $J_l$ and $K_l$ given as $(2l+1)\times (2l+1)$ matrices:
\begin{widetext}
\begin{eqnarray}
J_l = \left(\begin{array}{ccccc}
1-\frac{l}{n+1}c_2&\frac{\sqrt{l(n+1-l)}}{(n+1)^{3/2}}c_2&-\frac{\sqrt{ln(n+1-l)}}{(n+1)^{3/2}}c_2&\cdots& 0 \\
\frac{\sqrt{l(n+1-l)}}{(n+1)^{3/2}}c_2&1+\frac{l-nl-1}{(n+1)^2}c_2&\frac{(n+2-2l)\sqrt{n}}{(n+1)^2}c_2
&\cdots &0 \\
-\frac{\sqrt{ln(n+1-l)}}{(n+1)^{3/2}}c_2&\frac{(n+2-2l)\sqrt{n}}{(n+1)^2}c_2&1-c_1+\frac{(l-1)n+1-l-n^2}{(n+1)^2}c_2&
\cdots &0 \\
\vdots&\vdots&\vdots&\ddots &\vdots\\
0&0&0&\cdots&1-c_1+\frac{ln-n-n^2}{(n+1)^2}c_2\\
\end{array}\right),
\end{eqnarray}
\begin{eqnarray}
K_l = \left(\begin{array}{ccccc}
1-\frac{l}{n+1}c_2&\frac{\sqrt{l(n+1-l)}}{(n+1)^{3/2}}c_2&\frac{\sqrt{ln(n+1-l)}}{(n+1)^{3/2}}c_2&\cdots& 0 \\
\frac{\sqrt{l(n+1-l)}}{(n+1)^{3/2}}c_2&1+\frac{l-nl-1}{(n+1)^2}c_2&-\frac{(n+2-2l)\sqrt{n}}{(n+1)^2}c_2
&\cdots &0 \\
\frac{\sqrt{ln(n+1-l)}}{(n+1)^{3/2}}c_2&-\frac{(n+2-2l)\sqrt{n}}{(n+1)^2}c_2&1-c_1+\frac{(l-1)n+1-l-n^2}{(n+1)^2}c_2&
\cdots &0 \\
\vdots&\vdots&\vdots&\ddots &\vdots\\
0&0&0&\cdots&1-c_1+\frac{ln-n-n^2}{(n+1)^2}c_2\\
\end{array}\right).
\end{eqnarray}
\end{widetext}
All these real symmetric matrices can be diagonalized. To guarantee the non-negativity of 
$\Pi_0$, we only need to let all of their eigenvalues, which are real due to the symmetry in the above matrices, be non-negative. 
Through the induction on $n$, we obtained the eigenvalues of $J_l$ or $K_l$ in the diagonal elements of the following:
\begin{widetext}
\begin{eqnarray}
J_l^{\prime} = K_l^{\prime}=\left(\begin{array}{cccc}
\frac{2-c_1-c_2}{2}-\sqrt{\frac{c_1^2+c_2^2}{4}+\frac{(n^2-2n-1)c_1c_2}{2(n+1)^2}}&0&\cdots& 0 \\
0&\frac{2-c_1-c_2}{2}+\sqrt{\frac{c_1^2+c_2^2}{4}+\frac{(n^2-2n-1)c_1c_2}{2(n+1)^2}}&\cdots&0 \\
\vdots&\vdots&\ddots &\vdots\\
0&0&\cdots&1\\
\end{array}\right)\ .
\end{eqnarray}
\end{widetext}
The constant eigenvalue $1$ exists for all $J_l$'s and $K_l$'s, where $l=0,1,\cdots,n$. All the rest eigenvalues exist
in pairs, and the sum of each pair is $2-c_1-c_2$. After arranging the obtained eigenvalue pairs in all $J_l^{\prime}$'s and $K_l^{\prime}$'s 
by the members with the minus sign before the square
root in an ascending order, we see that
the least pair, $1-\frac{1}{2}c_1-\frac{1}{2}c_2\pm\sqrt{\frac{1}{4}c_1^2+\frac{1}{4}c_2^2+\frac{n^2-2n-1}{2(n+1)^2}c_1c_2}$, 
are the common eigenvalues of $J_l$ and $K_l$ 
for $l=1,2,\cdots,n$, and the second least,
 $1-\frac{1}{2}c_1-\frac{1}{2}c_2\pm\sqrt{\frac{1}{4}c_1^2+\frac{1}{4}c_2^2+\frac{n^2-6n+1}{2(n+1)^2}c_1c_2}$, the common
eigenvalues
of $J_l$ and $K_l$ for $l=2,3,\cdots.n$, and so on.
Therefore, if we require that the least common eigenvalue of all matrices except for $J_0$ and $K_0$ (the eigenvalue of which is the constant $1$) be non-negative, i.e.
\begin{eqnarray}
1-\frac{1}{2}c_1-\frac{1}{2}c_2-\sqrt{\frac{1}{4}c_1^2+\frac{1}{4}c_2^2+\frac{n^2-2n-1}{2(n+1)^2}c_1c_2}\geq 0\ ,\nonumber\\
\end{eqnarray}
the inconclusive operator $\Pi_0$ will be guaranteed to be non-negative.

Since $\langle\phi|\Pi_i|
\phi\rangle\leq 1$ for $\forall |\phi\rangle \in H$ and all $n$, we have $c_i\leq 1$ for $i=1,2$ and a non-negative number 
$1-\frac{1}{2}c_1-\frac{1}{2}c_2$. Then, from the above inequality, we express $c_2$ in terms of $c_1$:
\begin{eqnarray}
c_2\leq \frac{1-c_1}{1-\frac{2n+1}{(n+1)^2}c_1}\ .
\end{eqnarray}
To achieve the maximum success probability, we choose the equal sign, i.e. let the least eigenvalue of $\Pi_0$ be $0$.
Inserting the resulting expression into Eq. (15) gives
\begin{eqnarray}
\overline{P}=(\eta_1 c_1+\eta_2\frac{1-c_1}{1-\frac{2n+1}{(n+1)^2}c_1})\frac{n}{2(n+1)}\ .
\end{eqnarray}
We can easily find $c_{1,opt}$ where the right-hand side of this expression is maximized and the corresponding $c_{2,opt}$:
\begin{eqnarray}
c_{1,opt}&=&\frac{(n+1)^2}{2n+1}(1-\frac{n}{n+1}\sqrt{\frac{1-\eta_1}{\eta_1}}),\nonumber\\
c_{2,opt}&=&\frac{(n+1)^2}{2n+1}(1-\frac{n}{n+1}\sqrt{\frac{\eta_1}{1-\eta_1}}).
\end{eqnarray}
Substituting these optimum values into (15) gives the optimum result for POVM:
\begin{eqnarray}
\overline{P}_{POVM}(n,\eta_1)=\frac{n}{4n+2}\left(n+1-2n\sqrt{\eta_1(1-\eta_1)}\right ).~~~
\end{eqnarray}
Then we have the following optimum POVM elements:
\begin{eqnarray}
\Pi_{1, opt}(n,\eta_1) & = & \frac{(n+1)^2}{2n+1}(1-\frac{n}{n+1}\sqrt{\frac{1-\eta_1}{\eta_1}})\nonumber\\
&\times & (I_{E,T}-P_{E,T})\otimes I_O \nonumber\\
\Pi_{2, opt} (n,\eta_1)& = & \frac{(n+1)^2}{2n+1}(1-\frac{n}{n+1}\sqrt{\frac{\eta_1}{1-\eta_1}})\nonumber\\
&\times & (I_{O,T}-P_{O,T})\otimes I_E\ . 
\end{eqnarray}
Here we have assumed that {\it a prior} probabilities of the inputs satisfy $\eta_1 +\eta_2=1$, i.e. no failure in the preparation
period. 

\begin{figure}[t!]
\epsfig{file=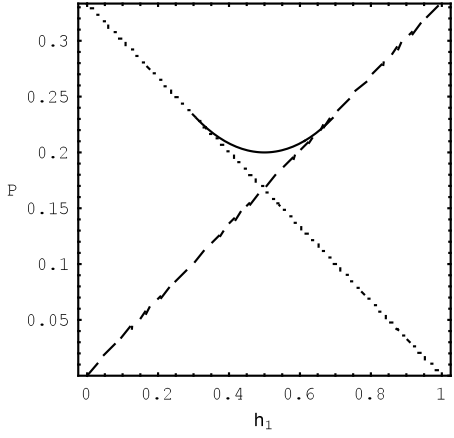, height=6cm}
\caption{Optimal average success probability, $P$, vs. the {\it a priori} 
probability, $\eta_{1}$, for $n=2$. Dashed line: $P_{1}$ from Eq. (31), dotted 
line: $P_{2}$ from Eq. (32), and solid curve: $P_{POVM}$ from Eq. 
(28). The optimal $P$ is given by 
$P_{2}$ for $\eta_{1}<4/13$, by $P_{POVM}$ for $4/13 \leq \eta_{1} \leq
9/13$ and by $P_{1}$ for $9/13<\eta_{1}$. The lower bound if the optimal average success probability reaches $0.2$.}
\label{F1}
\end{figure}

\begin{figure}[t!]
\epsfig{file=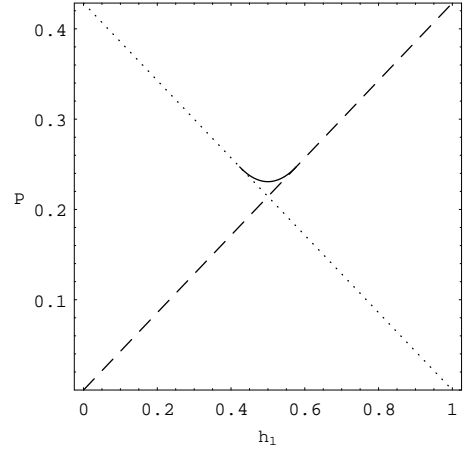, height=6cm}
\caption{Optimal average success probability, $P$, vs. the {\it a priori} 
probability, $\eta_{1}$, for $n=6$. The optimal $P$ is given by 
$P_{2}$ (dotted line) for $\eta_{1}<36/85$, by $P_{POVM}$ (solid curve) for $36/85 \leq \eta_{1} \leq
49/85$ and by $P_{1}$ (dashed line) for $49/85<\eta_{1}$. The lower bound if the optimal average success probability reaches $0.23$.}
\label{F2}
\end{figure}

This optimal POVM, however, holds valid only within the range of $\eta_1$ or $\eta_2$ where both 
$0\leq c_{1,opt}\leq 1$ and $0\leq c_{2,opt}\leq 1$ are satisfied simultaneously. From Eq. (27) we can find the range:
\begin{eqnarray}
\frac{n^2}{n^2+(n+1)^2}\leq \eta_1,\eta_2\leq \frac{(n+1)^2}{n^2+(n+1)^2}\ .
\end{eqnarray}
If $\eta_1=\frac{(n+1)^2}{n^2+(n+1)^2}$ ($\eta_2=\frac{n^2}{n^2+(n+1)^2}$), $\Pi_{1, opt}=(I_{E,T}-P_{E,T})\otimes I_O$ 
and $\Pi_{2, opt}=0$. 
The continuity to the outside of the POVM's validity domain requires this structure remain for 
$1\geq \eta_1 \geq \frac{(n+1)^2}{n^2+(n+1)^2}$. In other words, when the preparation is dominated by
the first input, the optimal measurement is realized by standard von Neumann measurement, which projects the input onto
the compliment of the totally symmetric subspace with respect to all the even digits in the program and the data digit.
Then, we get the success probability for each operator: $p_{1,opt}=\langle\Psi_{1}^{in}|(I_{E,T}-P_{E,T})\otimes I_O|\Psi_{1}^{in}\rangle$ 
and $p_{2,opt}=0$, indicating the second input is sacrificed completely. 
These results yield
the average success probability,
\begin{eqnarray}
\overline{P}_1(n,\eta_1)=\eta_1 \frac{n}{2(n+1)},
\end{eqnarray}
for $\eta_1 \geq \frac{(n+1)^2}{n^2+(n+1)^2}$.
Conversely, for $\eta_1=\frac{n^2}{n^2+(n+1)^2}$ ($\eta_2=\frac{(n+1)^2}{n^2+(n+1)^2}$), 
$\Pi_{2, opt}=(I_{O,T}-P_{O,T})\otimes I_E$ 
and $\Pi_{1, opt}=0$. Also from the continuity, we require this structure remain for 
$0\leq \eta_1 \leq \frac{n^2}{n^2+(n+1)^2}$.
This is a standard von Neumann measurement projecting the input onto
the compliment of the totally symmetric subspace with respect to all the odd digits in the program and the data digit.
Since the first input is sacrified completely in this case, we have
\begin{eqnarray}
\overline{P}_2(n,\eta_1)=\eta_2 \frac{n}{2(n+1)}=(1-\eta_1) \frac{n}{2(n+1)},
\end{eqnarray}
for $0\leq \eta_1 \leq \frac{n^2}{n^2+(n+1)^2}$. The situation is fully symmetric in the {\it a priori} 
probabilities of inputs. In the intermediate range, where neither of the input states dominates, the POVM does the job of
discrimination between them better than von Neumann measurement, while outside the range two full projectors do the work 
of identifying them at best.
The optimal average success probability of the device working in the whole range of {\it a priori} probability, 
as the generalization of the results in \cite{bergou}, is summarized as follows:
\begin{eqnarray}
&& \overline{P}^{opt}(n,\eta_1)\nonumber\\
&=&\left\{\begin{array}{cc}
\overline{P}_2(n,\eta_1) &             0\leq\eta_1< \frac{n^2}{n^2+(n+1)^2}, \\
\overline{P}_{POVM}(n,\eta_1) &      ~\frac{n^2}{n^2+(n+1)^2}\leq \eta_1 \leq \frac{(n+1)^2}{n^2+(n+1)^2}, \\
\overline{P}_1 (n,\eta_1)&  \frac{(n+1)^2}{n^2+(n+1)^2}< \eta_1\leq 1 . 
\end{array}
\right. \nonumber\\ 
\end{eqnarray} 
It is seen from the above results that the optimal success probability changes continuously at the boundaries of the respective 
regions of validity for the three measurements.

\begin{figure}[ht]
\epsfig{file=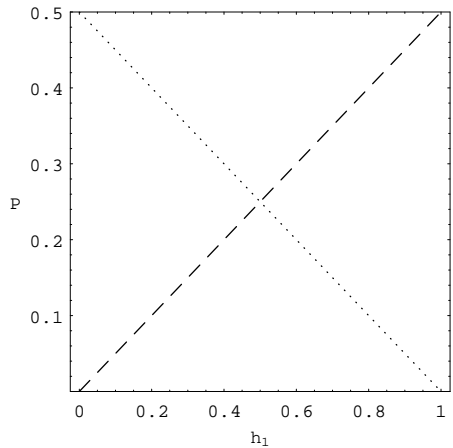, height=6cm}
\caption{Optimal average success probability, $P$, vs. the {\it a priori} 
probability, $\eta_{1}$, for $n=\infty$. In this upper bound limit, the optimal scheme reduces to two full projector probabilities 
$P_{1}$ (dashed line) and $P_{2}$ (dotted line). The minimum value can reach $0.25$.}
\label{F3}
\end{figure}

Some features should be noticed for this device: 
First, from Eq. (30), the validity range of the POVM measurement 
will become narrower and narrower with the increase of $n$, the number of copies
 used in the program. As $n$ tends to infinity, this range will shrink to a point at $\eta_1=\eta_2=0.5$. It means
that the optimum measurement will reduce to only von Neumann in this extreme. Second, if the preparation probability is fixed,
the success probability of the device will increase with the number of copies of the states stored in the program.
The larger size of the program yields more information about the unknown states.
For example, at $\eta_1=\eta_2=0.5$, where the identification of the input state is the hardest and the difference between the POVM and two
full projectors is the largest, the average success probability, $\overline{P}_{POVM}(n)= n/(4n+2)$, increases from $1/6$ to $1/4$ by $50\%$ as $n$ goes from $1$ to infinity. We depict these features in Figs. 1-3 for three different 
$n$ ($n=2,6,$ and $\infty$).

Another interesting feature is that the optimal measurement operators and the validity domain depend only on
the number of the states we put into the program and their preparation probabilities. Since they are independent of
the information encoded in a specific pair of unknown qubits $|\psi_1\rangle$ and $|\psi_2\rangle$, the device performs 
universally. It can be programed with whatever couple of 
unknown qubits we want to identify and discriminates between the produced registers with the best chance of success 
and no error. 

Finally, we compare the obtained results with those of UD for the averages of systems. Suppose the {\it a prior}
probabilities of the systems prepared as in Eq. (1) are equal. If we take the averages of the density matrices 
$\rho_1=|\Psi_{1}^{in}\rangle\langle\Psi_{1}^{in}|$ and $\rho_2=|\Psi_{2}^{in}\rangle\langle\Psi_{2}^{in}|$ over the
Bloch spheres of the uniformly and independently distributed $|\psi_1\rangle$, $|\psi_2\rangle$ and find the 
optimal UD solution to these two averages, which are two uniformly distributed {\it known} mixed states, we will have the optimal 
average success probability of $1/3$ if the number of the qubit copies tends to infinity \cite{hayashi,he1}. This is equal 
to the average of IDP limit [3]-[5] for discriminating a pair of known qubits. The corresponding maximum average success 
probability, $1/4$, of our device is obtained through first measuring the systems with the optimal POVM and then 
taking the averages of the expectation values of $\Pi_1$ and $\Pi_2$, and is lower than this IDP average. 
However, the device described in the current work realizes a practical identification of any pair of {\it randomly} distributed unknown qubits instead of their averaged input states. We see from this fact that the discrimination of two unknown systems (with only some 
information about their symmetry available) and the discrimination of their known averages are not equivalent. 

So far we have worked out a scheme for a device that unambiguously discriminate between pairs of quantum registers 
produced with more than one copy of unknown qubits used in the program or reference digits. Its maximum 
average success probability increases with the number of the copies we store in the program, and also depends on the {\it a prior} probability of the states to be discriminated. 
We can tune the parameters $c_1$ and $c_2$ of the setup according to the copy number $n$ and the 
{\it a prior} probability $\eta_1$ or $\eta_2$, so that it optimally performs in the required measurement.

The authors would like to thank M. Hillery for valuable discussions.

\end{document}